\documentclass{ws-procs975x65}

\begin{document}

\def\vs{\vspace{1.0cm}}

\def\lsim{\mathrel{\vcenter{\hbox{$<$}\nointerlineskip\hbox{$\sim$}}}}
\def\gsim{\mathrel{\vcenter{\hbox{$>$}\nointerlineskip\hbox{$\sim$}}}}
\def\ol{\overline}
\def\ul{\underline}
\def\req#1{(\ref{#1})}
\def\eq#1{Eq.~(\ref{#1})}
\newcommand{\ba}[1]{\begin{eqnarray} \label{(#1)}}
\newcommand{\ea}{\end{eqnarray}}
\newcommand{\rf}[1]{(\ref{(#1)})}

\def\half{\frac{1}{2}}
\def\eps{\epsilon}
\def\jth{$j^{\rm th}$}
\def\ev {\ \mbox {eV}}
\def\kev{\ \mbox{KeV}}
\def\mev{\ \mbox{MeV}}
\def\gev{\ \mbox{GeV}}
\def\tev{\ \mbox{TeV}}
\def\egzk{E_{\rm GZK}}
\def\dgzk{D_{\rm GZK}}
\def\tentwenty{10^{20}}
\def\mjbar{\ol{m_j}}
\def\dmpc{D_{\rm Mpc}}
\def\lammpc{\lambda_{\rm Mpc}}
\def\nua{\nu_{\alpha}}
\def\nub{\nu_{\beta}}
\def\nue{\nu_e}
\def\nunote{\nu_{\not e}}
\def\numu{\nu_{\mu}}
\def\nutau{\nu_{\tau}}
\def\nuebar{{\bar \nu_e}}
\def\numubar{{\bar \nu_\mu}}
\def\nus{\nu_s}
\def\nuetilde{\tilde{\nu}_e}
\def\numutilde{\tilde{\nu}_{\mu}}
\def\nutautilde{\tilde{\nu}_{\tau}}
\def\dmsq{\delta m^2}
\def\dmatm{\delta m^2_{\rm atm}}
\def\dmsun{\delta m^2_{\rm sun}}
\def\dmlsnd{\delta m^2_{\rm LSND}}

\title{Physics with Cosmic Neutrinos, PeV to ZeV}

\author{Thomas J. Weiler}

\address{ Department of Physics \& Astronomy, Vanderbilt University,
Nashville, TN 37235 USA\\
E-mail: tom.weiler@vanderbilt.edu}

\maketitle

\abstracts{
Neutrinos offer a particularly promising eye on the extreme Universe.
Neutrinos are not attenuated by intervening radiation fields such as the
Cosmic Microwave Background, and so 
they are messengers from the very distant and very young phase of
the universe.
Also, neutrinos are not deflected by cosmic magnetic fields, and so
they should point to their sources.
In addition, there are particle physics aspects of neutrinos 
which can be tested only with cosmic neutrino beams.
After a brief overview of highest-energy cosmic ray data,
and the present and proposed experiments 
which will perform neutrino astronomy,
we discuss two particle physics aspects of neutrinos.
They are possible long-lifetime decay of the neutrino,
and a measurement of the neutrino-nucleon cross-section at a CMS energy 
orders of magnitude beyond what can be achieved with terrestrial accelerators.
Measurement of an anomalously large neutrino cross-section
would indicate new physics
(e.g. low string-scale, extra dimensions, precocious unification),
while a smaller than expected cross-section
would reveal an aspect of QCD evolution.
We then discuss aspects of neutrino-primary models for the extreme-energy (EE)
cosmic ray data.  Primary neutrinos in extant data 
are motivated by the directional clustering at EE 
reported by the AGASA experiment.
We discuss 
the impact of the strongly-interacting neutrino hypothesis on lower-energy 
physics via dispersion relations, 
the statistical significance of AGASA directional clustering,
and the possible relevance of the Z-burst mechanism 
for existing EE cosmic ray data.
}

\section{Introduction}
\label{sec:intro}
Detection of ultrahigh-energy neutrinos is one of the important
challenges of the next generation of cosmic ray detectors.  Their discovery
will mark the advent of neutrino astronomy, allowing the mapping on the
sky of the most energetic, and most distant, sources in the Universe.  In
addition, detection of extreme-energy (EE) neutrinos,
those at $10^{20}$~eV and beyond, may help resolve 
puzzles associated with the giant air-shower events
observed with energies beyond the
Greisen-Zatsepin-Kuzmin (GZK) limit of 
$E_{\rm GZK}\equiv 5\times 10^{19}$~eV.
In this context, neutrino observations may validate 
Z-bursts, topological defects, superheavy
relic particles, new strong-interactions, {\it etc.}

Measurement of the neutrino cross-section itself at EE
is of considerable importance to particle physics.
Our highest-energy knowledge of the neutrino-nucleon cross-section 
comes from the HERA experiment, at a CMS energy of 0.2~TeV.
The CMS energy for a CR of energy $E=10^{20}\,E_{20}$~eV on an 
air nucleus is 
$\sqrt{s}=0.5\,\sqrt{E_{20}}$~PeV.
This can be compared to the values of terrestrial hadron accelerators,
2~TeV at Fermilab, and 14~TeV to occur at CERN's LHC.
Furthermore, if the primary CR is a neutrino, its energy is 
not shared among partons, so its ``reach'' is even larger than the
comparison with Fermilab and the LHC indicates.
Estimates of the cross-section at $10^{20}$~eV 
require QCD extrapolations over three orders of magnitude
beyond HERA in CMS energy.  Thus, a measurement of the cross-section at 
$10^{20}$~eV will test QCD with a large lever arm.
Additionally, there may well be new non-SM physics 
revealing itself in the neutrino sector at EE,
arising from 
new physics thresholds not accessible to terrestrial accelerators.
Needless to say, the value of the cross-section is also crucial 
to the future evolution of ``neutrino telescope'' detectors,
for the neutrino event rate
is proportional to this cross-section (more on this later).

And finally, we note that the stability of the neutrino is 
best tested with PeV cosmic neutrinos.
The ``cosmic'' aspect guarantees a long decay path,
while the PeV energy is a compromise between
minimizing the boost factor $\gamma =E/m$ and ensuring
events from extragalactic rather than atmospheric sources.


Several sources of neutrinos with PeV to ZeV ($10^{21}\;{\rm eV}$) 
energies are possible,
ranging conservatively from AGNs and GRBs,
to exotic top-down production.
The latter may in principle even provide energies up to 
the grand-unified mass of $\sim 10^{24}$~eV.
A nice review of sources, classified according to their speculative nature,
was given a few years ago by Protheroe.~\cite{Protheroe} 
Some of the most recent ideas for EE neutrino sources 
are discussed in our later section \ref{sec:Zbursts} 
on Z-bursts.

In addition to the flux of neutrinos produced in 
cosmic engines, there is also a reasonably guaranteed
prediction for a flux 
of ``GZK neutrinos''
in the energy range $10^{15}$ to $10^{20}$~eV, based on
the observed flux of cosmic ray (CR) protons at and above 
the GZK limit.
These neutrinos result from the decay of charged
pions photo-produced by the interaction of super-GZK nucleons 
on the CMB background. 
This flux is expected to peak in the decade
$10^{17}$ to $10^{18}$~eV for uniformly-distributed
proton sources, and around $10^{19}$~eV for ``local'' sources
within $\sim 50$~Mpc of earth. \cite{ESS01}

On the experimental side, 
past observations of EE cosmic rays have 
been made by the Volcano Ranch, Haverah Park, 
AKENO, and Fly's Eye experiments. \cite{NagWat}
The record energy, held by the Fly's Eye event of a decade ago,
is $3\times 10^{20}$~eV, or 50~Joules.
This truly macroscopic amount of energy equals that 
in a professional baseball
pitcher's fastball, which consists of $10^{27}$ nucleons.  
This latter comment shows that somehow Nature is $10^{27}$
times more efficient at acceleration than is the human arm.

\begin{figure}[th]
\epsfxsize=12cm
\epsfbox{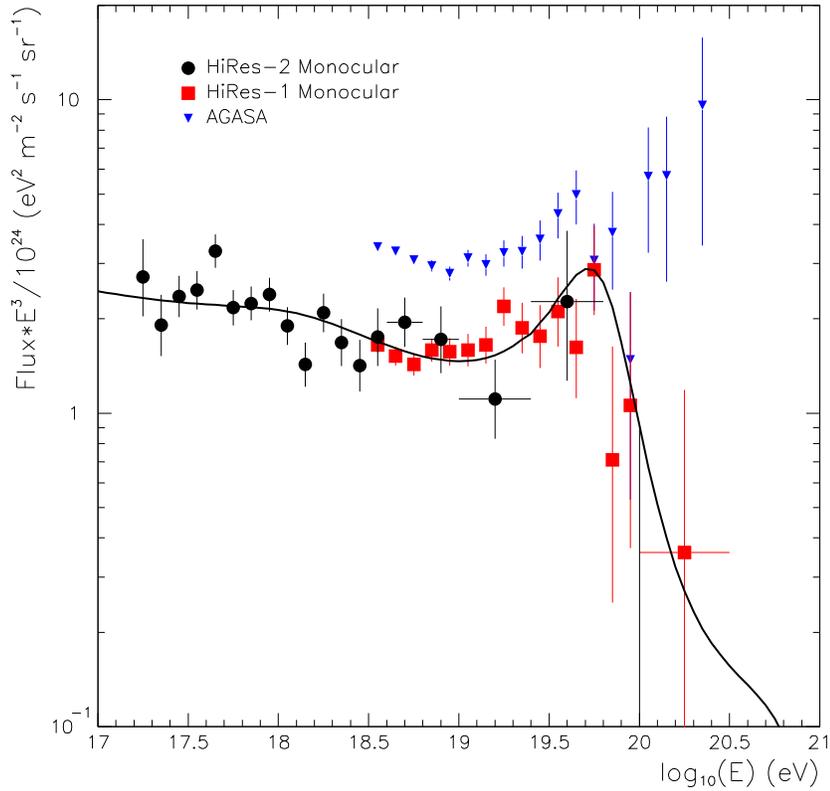} 
\caption
{$E^{3}$ times the UHE Cosmic Ray Flux, with data from HiRes and AGASA
(from T. Abu-Zayyad et al., HiRes Collaboration 
Also shown is a fit to the HiRes data of a uniform cosmic distribution 
of sources cutoff at $E=10^{21}$~eV and subject to $e^+ e^-$ losses and  
the GZK photo-pion absorption mechanism.  
Not shown here is the Fly's Eye event at $3\times 10^{20}$~eV. 
\label{fig:AvsH} }
\end{figure}

Most of the catalogued EE events come from the AGASA experiment,
the successor to AKENO.  The HiRes experiment, the successor 
to Fly's Eye, has an exposure similar to AGASA, but an EE event rate
lower by almost an order of magnitude.
The discrepancy between these two experiments is 
shown in Fig.\ \ref{fig:AvsH}.
There is clearly a systematic error in the energy assignments of 
(at least) one of the experiments.  
It has been noted, and is somewhat evident in the figure,
that even the spectral dip (``ankle'') occurs at different energies
in the two experiments.
The AGASA experiment uses ground-based scintillator to measure showers,
whereas HiRes uses optics to measure the near-UV light emitted by 
atmospheric $N_2$ excited by the passing shower.
Fortunately, both measurement techniques will be present in the 
much larger AUGER experiment under construction in the high plateau
of Argentina.  In this experiment,
simultaneous measurement of the same shower with both techniques
will allow a cross-check of systematics effects,
thereby settling in the next two years
the issue of AGASA vs.\ HiRes.
De Marco et al.\ \cite{DBO} 
give a quantitative appraisal of the AGASA vs.\ HiRes 
discrepancy, and the resolving potential of AUGER, and ultimately, of
the ``Extreme Universe Space Observatory'' (EUSO) experiment.~\cite{EUSO} 
It is worth noting that if the lower HiRes rate turns out to be correct,
and even if the GZK cutoff is realized,
the region above $E_{\rm GZK}$ remains a fertile arena for new physics.
The depletion or total absence of nucleon primaries in this energy 
region would constitute a background free playground for the discovery
of new primary particles, such as neutrinos, 
supersymmetric particles,\cite{LSPflux}
magnetic monopoles,\cite{monopoles} and any other exotic quanta.
Moreover, the MFP for photons above $10^{21}$ rapidly increases to  
$\sim 100$~Mpc.  
This means that if there are sources of such energetic photons,
they should be unmasked by the proposed larger aperture experiments.

The EUSO experiment will mount a two meter Fresnel lens on the 
International Space Station (ISS), 400~km above the earth.  
The lens will look down
on the earth's atmosphere, and focus the near-UV $N_2$ fluorescence
emitted in  air-shower cascades occurring in the moonless night sky.
EUSO is scheduled for deployment in early 2009.
Eventually, the free-flying multi-satellite 
experiment ``OWL'' may follow EUSO.
In terms of US states, the Auger field of view
may eventually equal the area of Rhode Island,
while EUSO and OWL may equal Texas, or more.
The target-mass of atmosphere available to EUSO is a teraton.
For these space-based experiments, 
the $1/r^2$ loss at the 400-500 km height limits
their sensitivity to events with energy above $\sim 10^{19}$~eV,
thereby providing a natural filter to select only
the most extreme-energy CRs.

The ICECUBE experiment is optimized to measure the 
neutrino flux at a PeV and above.
It consists of strings of optical modules triggering on the 
Cerenkov light emitted by showers and muons in a gigaton 
of Antarctic Ice.  
This experiment builds on AMANDA, the proof-of-principle prototype experiment
presently operating in South Polar ice.
ICECUBE is funded, and is expected to be fully arrayed 
near the end of this decade.
Prototypes for ocean detectors are also under construction.
These are ANTARES and NESTOR in the Mediterranean Sea.
A proposed gigaton array is named NEMO, possibly to be sited off of Sicily.

The far future of EE neutrino astronomy 
may belong to radio frequency detectors.
The idea here is that the amplitude for Cerenkov emission 
at a particular wavelength $\lambda$ 
coherently sums the net charge of the shower in the length $\lambda$.
Since the net charge is known to be about 20\% of the total charge,
and the total charge scales simply with shower energy as $\sim E/{\rm GeV}$,
it follows that the Cerenkov rate scales as $(E/{\rm GeV})^2$ 
at long wavelength.
Thus, radio wavelengths are tremendously favored at extreme-energy.
The RICE experiment at the South Pole is a prototype for 
radio detection.
The experiment ANITA may be the first to witness an EE cosmic neutrino event.
ANITA is a balloon experiment at the South Pole, 
configured to detect radio signals generated by showers
originating from ``earth-skimming'' neutrinos 
interacting just below the horizontal
(more on earth-skimmers in section \ref{sec:cantlose}).
Eventually, large radio arrays may be deployed with ICECUBE, 
or in the massive salt domes which are known to populate 
the underground in the southeastern US.

Another noteworthy experimental idea is the 
mountain-to-mountain ``telescope'' configured to 
see horizontal, neutrino-induced showers emanating from one mountain
toward the lens located across the valley on the other.~\cite{mtn}
A useful table of many happening and proposed EE neutrino experiments, 
assembled by Peter Gorham, is available at \\
$http://astro.uchicago.edu/home/web/olinto/aspen/gorham\_table.htm$.

In what follows we develop each of the theoretical themes noted 
earlier in this introduction.

\section{Astrophysical Test of Neutrino Decay}
\label{sec:DK}

Neutrinos from astrophysical sources are expected to arise dominantly
from the decays of pions and their muon daughters, which results in
initial flavor ratios
$\phi_{\nu_e}:\phi_{\nu_{\mu}}:\phi_{\nu_{\tau}}$ of nearly $1:2:0$.
As the cosmic neutrinos travel over many oscillation lengths,
they lose their phase information (``decohere'') and arrive at earth 
as an incoherent ensemble of mass eigenstates.
The relative fluxes of each mass eigenstate are given by $\phi_{\nu_i}=
\sum_{\alpha} \phi_{\nu_\alpha}^{\rm source} |U_{\alpha i}|^2$, where
$U_{\alpha i}$ are elements of the neutrino mixing matrix.  
For three active neutrino species there is now strong
evidence to suggest that $\nu_{\mu}$ and $\nu_{\tau}$ are maximally
mixed and $U_{e3} \simeq 0$.  Consequently,
the neutrino mass eigenstates are
produced in the ratios 
$\phi_{\nu_1}:\phi_{\nu_2}:\phi_{\nu_3} = 1:1:1$,
and so arrive at earth in this same ratio.
Due to the newly determined values of neutrino mixing parameters 
with their small observational uncertainties, this ratio is robust.

It was recently shown\cite{BBHPW} 
that this robustness allows for  
sensitive tests of neutrino decay, as decay would alter the measured
flavor ratios in a strong and distinctive
fashion.  Conversely, a measurement of ratios other than 1:1:1 
can be construed as evidence for
unstable neutrino mass.
It was further shown that
neutrino decay cannot be mimicked by either different
neutrino flavor ratios at the source or other non-standard neutrino
interactions.  In this section we elaborate on these results.

We restrict our attention to the two body decays
\begin{equation}
\nu_i \rightarrow \nu_j + X \;\;\; \rm{and} \;\;\;
\nu_i \rightarrow \overline{\nu}_j + X,
\end{equation}
for which limits are too weak to eliminate the
possibility of astrophysical neutrino decay by a factor of about $10^7
\times (L/100 {\rm\ Mpc}) \times (10 {\rm\ TeV}/E)$.~\cite{BB}
Here, $\nu_i$ are neutrino mass eigenstates and $X$ denotes a very
light or massless particle, e.g. a singlet Majoron.  
Radiative two-body decay modes and three-body decays 
of the form $\nu \rightarrow \nu\nu\bar{\nu}$ need not be considered,
as they are already constrained beyond what can be inferred from 
cosmic fluxes,
by the absence of photons in appearance searches,
and by bounds on anomalous
$Z\nu\bar{\nu}$ couplings, respectively.  

We also assume that the decays are complete, i.e. 
none of the decaying state arrives at earth.
This is reasonable because even the shortest distances 
are typically hundreds of Mpc, and
the typical energies in a steeply falling spectrum are not too large.
The assumption of complete
decay means that the distance and intensity
distributions of sources need not be considered.  
Finally, neutrinos and antineutrinos need not be  
distinguished because their cross sections
rapidly approach each other above 10 TeV.

First suppose that there are no detectable decay products, that is, 
the decaying neutrinos simply disappear.  
Such would be the case for decay to
``invisible'' daughters such as a sterile neutrino, or for decay to 
active daughters if the source spectrum falls sufficiently steeply with 
energy so that the flux of daughters with degraded energy 
makes a negligible contribution to the total flux at that energy.
Since coherence is lost, one has for the flavor fluxes
(assuming completeness: $L\gg \tau_i$),
\be
\label{simple}
\phi_{\nu_\alpha}(E) \longrightarrow
\sum_{s, \beta} \phi^{\rm source}_{\nu_\beta}(E)
|U_{\beta s}|^2 |U_{\alpha s}|^2,
\ee
where the sum on $s$ is over just the stable states.

The simplest case (and the most generic expectation) is a normal
hierarchy in which both $\nu_3$ and $\nu_2$ decay, leaving only the
lightest stable eigenstate $\nu_1$.  In this case the flavor ratio is
$U_{e1}^2:U_{\mu1}^2:U_{\tau1}^2$.  Neglecting $U_{e3}=0$, one then has
\begin{equation} 
\label{generic}
\phi_{\nu_e}:\phi_{\nu_\mu}:\phi_{\nu_\tau} =
\cos^2 \theta_{\odot}:\frac{1}{2}\sin^2 \theta_{\odot}:
\frac{1}{2}\sin^2 \theta_{\odot} \simeq 6:1:1,
\end{equation}
where $\theta_{\odot}$ is the solar neutrino mixing angle, which we
set to $30^{\circ}$.   
In the case of an inverted hierarchy, $\nu_3$ is the lightest and
hence stable state, and so
\begin{equation}
\label{inverted}
\phi_{\nu_e}:\phi_{\nu_\mu}:\phi_{\nu_\tau} =
U_{e3}^2:U_{\mu3}^2:U_{\tau3}^2 = 0:1:1.
\end{equation}

Interestingly, both cases have extreme deviations of
the flavor ratio from the 1:1:1 in the absence of decays,
which provides a very useful diagnostic.  
Assuming no new physics besides decay, a ratio of 
$\phi_{\nu_e}:\phi_{\nu_\mu}$
greater than 1 suggests the normal hierarchy, while a ratio
smaller than 1 suggests an inverted hierarchy.   It is also interesting
to note that complete decay cannot reproduce $1:1:1$.  One of the mass
eigenstates does have a flavor ratio similar to $1:1:1$, but it is the
heavier of the two solar states and cannot be the lightest,
stable state.  (A possible but unnatural exception occurs if only this
state decays).

These clear and striking predictions depend strongly on the recent progress 
which determined the neutrino mixing parameters.  
In particular, it is very
significant that $\theta_\odot \simeq 30^\circ$ is well
below the maximal $45^\circ$, for which Eq.~(\ref{generic}) would
instead be a much less dramatic $2:1:1$.  In addition, $\theta_\odot <
45^\circ$ means that $\delta m^2_{12} > 0$ and hence that $\nu_2$
(with flavor ratios $0.7:1:1$) can never be the lightest mass
eigenstate.  Maximal $\theta_{\rm atm}$ and very small
$U_{e3}$ also make the predictions clearer.  

Quite different ratios may result,
depending on which of the mass eigenstates are unstable, the decay
branching ratios, and the hierarchy of the neutrino mass eigenstates.
Also, when the appearance of daughter neutrinos from the 
decay cannot be neglected,
the equations are more complicated, but only slightly so.
That case is also treated in Beacom et al,~\cite{BBHPW}
for both Dirac and for Majorana neutrinos.
I do not treat that case here, but 
instead, list some possibilities for the normal hierarchy in Table~I.
\begin{table}[th]
\label{table:ratios}
\caption{Flavor ratios for various decay scenarios
(normal hierarchy only).}
\begin{center}
\begin{tabular}{|c|l|l|c|}
\hline\hline
Unstable & Daughters & Branchings &  
$\phi_{\nu_e}:\phi_{\nu_\mu}:\phi_{\nu_\tau}$ \\
\hline\hline
$\nu_2$, $\nu_3$  & anything & irrelevant & $6:1:1$ \\
\hline
$\nu_3$	          & sterile  & irrelevant & $2:1:1$ \\
\hline
$\nu_3$          & full energy & $B_{3 \rightarrow 2}=1$  & $1.4:1:1$ \\
                 & degraded ($\alpha=2$)               &  & $1.6:1:1$ \\
\hline
$\nu_3$          & full energy & $B_{3 \rightarrow 1}=1$   & $2.8:1:1$ \\  
                 & degraded ($\alpha=2$)               &   & $2.4:1:1$ \\
\hline
$\nu_3$          & anything & $B_{3 \rightarrow 1}=0.5$    & $2:1:1$ \\
                 &  &  $B_{3 \rightarrow 2}=0.5$           & \\  
\hline\hline  
\end{tabular}
\end{center}
\end{table}

An important issue is whether there are
other scenarios (either non-standard astrophysics or
non-standard neutrino properties) that would give similar ratios.
The answer is that 
since the mixing angles $\theta_\odot$ and $\theta_{\rm atm}$ 
are both large, and since the neutrinos are produced and detected in 
flavor states, no initial flavor ratio can result in a measured 
$\phi_{\nu_e}:\phi_{\nu_\mu}$ ratio anything like that of our two main cases, 
$6:1$ and $0:1$.
In terms of non-standard particle physics, decay is unique in the sense 
that it is ``one-way''. 
Oscillations or magnetic moment
transitions are, on the other hand, ``two-way''.  
Since the initial flux ratio in the mass basis is $1:1:1$, 
magnetic moment transitions between (Majorana) mass eigenstates cannot 
alter this ratio, due to the symmetry between $i\rightarrow j$ and 
$j \rightarrow i$ transitions.
However, if neutrinos have Dirac masses, magnetic moment transitions
(both diagonal and off-diagonal) turn active neutrinos into sterile states, so
the same symmetry is not present.  Yet the process will 
average out at 1/2, so there is no way 
to leave only a single mass eigenstate.

Experimentally,
the number of muon tracks and the number of
showers (charged- and neutral-current combined) are accessible.
The relative number of shower events to track events can be related to the
most interesting quantity for testing decay scenarios, i.e., the $\nu_e$
to $\nu_\mu$ ratio.  The precision of the upcoming experiments, e.g. 
ICECUBE, should be
good enough to test the extreme flavor ratios produced by decays,
and thereby discover or limit neutrino decay.

\section{Can't Lose Theorem for Smaller/Larger Neutrino Cross-Section}
\label{sec:cantlose}

The expected rates for neutrino observation 
in approved and proposed experiments
are proportional to $\sigma_{\nu N}$.
The grossly-extrapolated cross section at $10^{20}$~eV is $\sim 10^{-31}{\rm cm}^2$.
If Nature offers a smaller cross section, then 
the main detection signal proposed for
UHE neutrino experiments would be compromised.
On the other hand, the extrapolated cross-section may be too low,
for it ignores possible contributions from new physics that may enter in the
$\sim$TeV to PeV scale accessible to CR physics but 
inaccessible to terrestrial accelerators.

\begin{figure}[th]
\centerline{\epsfxsize=12cm\epsfbox{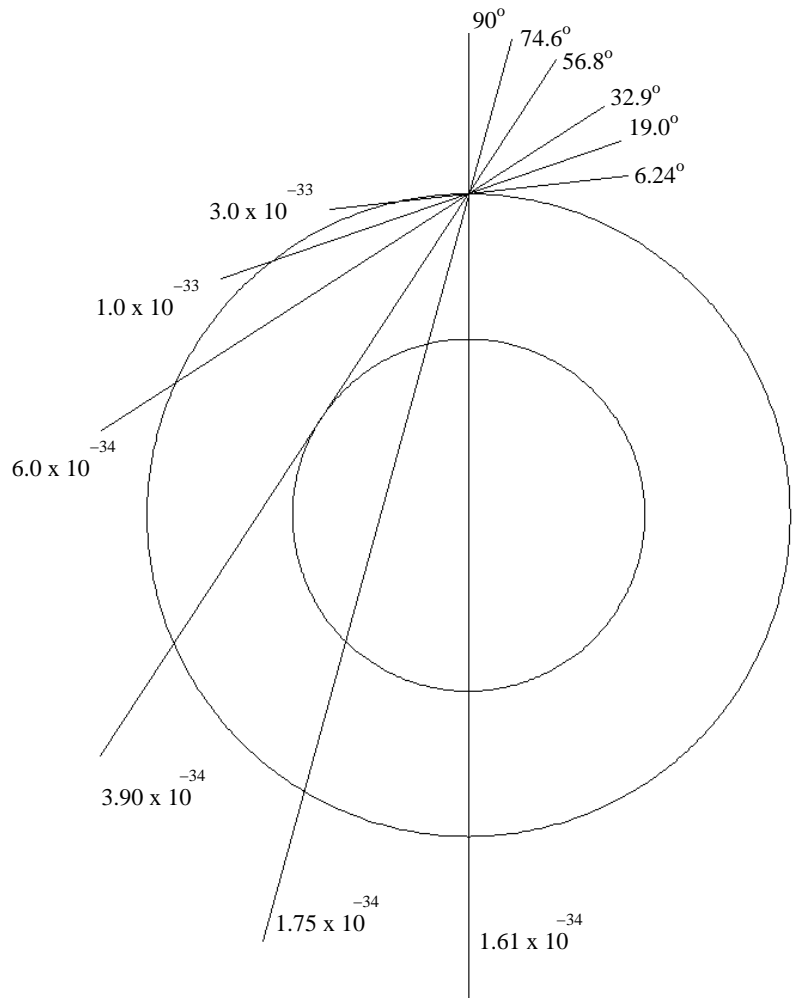}}
\caption{Neutrino MFP displayed as a chord length in the earth.
Cross-section values label the various MFP's.
\label{fig:NuEarthMFP}}
\end{figure}

\begin{figure}[th]
\centerline{\epsfxsize=12cm\epsfbox{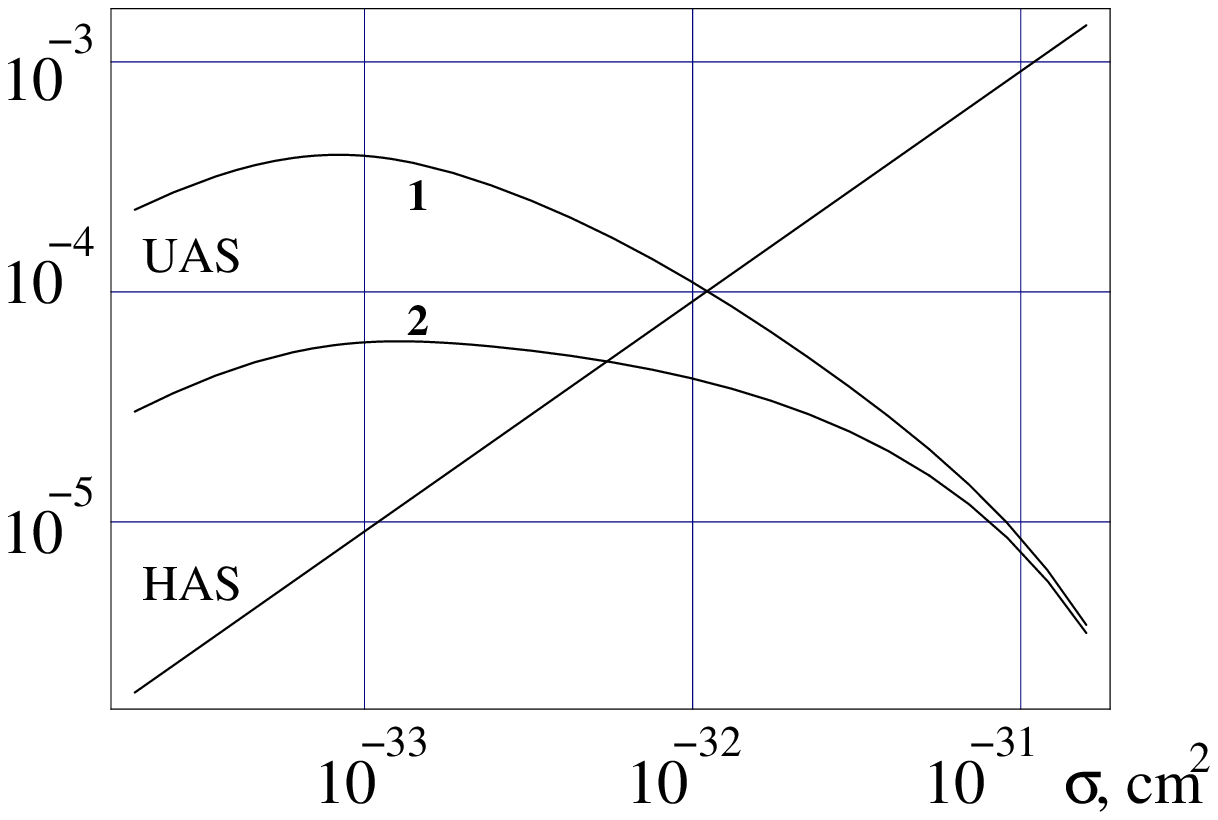}}
\label{fig2}
  \caption{The air shower probability per incident tau
neutrino $R_{\rm UAS}/F_{\nu_\tau} \pi A$ as a function of the neutrino
cross section (eq.(7)).
The incident neutrino energy is $10^{20}$~eV and the assumed
energy threshold for detection of UAS is
$E_{\rm th}=10^{18}$eV for curve 1 and $ 10^{19}$eV for curve 2.
}
\end{figure}

Consider, for example, the space-based experiments EUSO and OWL.
The event rate for nearly horizontal air
showers (HAS) resulting from $\nu$-air
interactions in the Earth's atmosphere 
is proportional to $\sigma_{\nu N}$.
Fortuitously, it was recently shown\cite{KWanysig}
that the flux of up-going
charged leptons (UCL) per unit surface area
produced by neutrino interactions {\sl below} the Earth's 
surface is {\sl inversely} proportional to $\sigma_{\nu N}$,
as long as the neutrino absorption mean free path in Earth
is small in comparison with the Earth's radius ($R_\oplus$);
i.e. for $\sigma_{\nu N}  \stackrel{>}{_{\scriptstyle \sim}}
 2\times 10^{-33}{\rm cm}^2$ (see Fig.\ \ref{fig:NuEarthMFP}).
This contrasts with the HAS rate 
proportional to $\sigma_{\nu N}$.
Even when showering of the upgoing tau lepton is included,
the upgoing air-shower (UAS) rate still varies inversely
with the cross-section in the range of interest, and can even 
exceed the HAS rate by several orders of magnitude,
as displayed in Fig.~\ref{fig2}.
%
%
%

Taken together, up-going and horizontal rates ensure a
healthy total event rate,
regardless of the value of $\sigma_{\nu N}$.
Moreover, by comparing the HAS
and UAS rates, the neutrino-nucleon cross section can
be inferred at energies as high as $10^{11}$~GeV or higher.
This enables QCD studies at a minimum,
and possibly discovery of a strong neutrino cross-section.
$\sigma_{\nu N}$ may also be determinable from
a measurement of the angular distribution of UCL/UAS
events, in addition to the approach comparing UAS and HAS rates.
One expects the angular distribution of UCL to peak near
$\cos\theta_{\rm peak}\sim
\left( 2\,\langle\rho\rangle\,R_\oplus\,\sigma_{\nu N}\right)^{-1}\,.$

Let us now examine the physics of upward showers in some detail.
UHE neutrinos are expected to arise from pion and subsequent muon decay.
These flavors
oscillate and eventually decohere during their Hubble-time journey.
If $|U_{\tau 3}|\simeq |U_{\mu 3}|$ are large, as inferred from
the Super-Kamiokande data, then
$F_{\nu_\tau}\approx\frac{1}{3}F_\nu$ is expected.
Because the energy-loss MFP for a $\tau$ produced in rock or
water is much longer than that of a muon or electron, 
the produced taus have a much
higher probability to emerge from the Earth and to produce an atmospheric
shower.  Thus, the
dominant primary for initiation of UAS 
``earth-skimming'' events is the tau neutrino.~\cite{FFWY} 

Consider an incident tau neutrino whose trajectory
cuts a chord of length $l$ in the Earth.
The probability for this neutrino to reach a
distance $x$ is $P_\nu (x) =e^{-x/\lambda_\nu}$, where
$\lambda_\nu^{-1}=\sigma_{\nu N}\,\rho$
(the conversion from matter density to number density via
$N_A/{\rm gm}$ is implicit).
The probability to produce a tau lepton in the interval $dx$ is
$\frac{dx}{\lambda_\nu}$.  
The $\tau$
produced at point $x$ emerges from the surface with energy
$E_\tau\sim E_\nu\,e^{-(l-x)/\lambda_\tau}$.
To produce an observable shower, one requires 
$P_{\tau \rightarrow {\rm UAS}}=\Theta (\lambda_\tau+x-l)$, with
$\lambda_\tau=(\beta_\tau\,\rho_{sr})^{-1}\,\ln(E_\nu/E_{\rm th})$;
$\beta_\tau \approx 0.8 \times 10^{-6} {\rm cm}^2/g$
is the exponential energy-attenuation coefficient
and $E_{\rm th}$ is the minimum detectable energy.
Taking the product of these conditional probabilities and integrating over
the interaction site $x$ we get the probability for a tau neutrino
incident along a chord of length $l$ to produce an UCL:
\begin{equation}
\label{probUCL}
P_{{\nu_\tau}\rightarrow\tau}(l)= \int_{l-\lambda_\tau}^l
\frac{dx}{\lambda_\nu} \, e^{-x/\lambda_\nu}
=(e^{\lambda_\tau/\lambda_\nu}-1) \,e^{-l/\lambda_\nu}\,.
\end{equation}
The emerging tau decays in the atmosphere with probability\\
$P_d= 1-\exp(-2 R_\oplus H/c\tau_\tau l) $, where $H\approx 10$~km
parametrizes the height of the atmosphere.
Thus, the probability for a tau-neutrino to produce an
UAS is
\begin{equation}
\label{probUAS}
P_{{\nu_\tau}\rightarrow{\rm UAS}}(l)=
( 1-e^{-2 R_\oplus H/c\tau_\tau l} )
\, P_{{\nu_\tau}\rightarrow\tau}(l)\,.
\end{equation}
The fraction of neutrinos with chord lengths in the
interval $\{l, l+dl\}$ is $\frac{l}{2 R_\oplus^2} dl$.
Finally, including two further geometric factors,
the solid angle $\pi$ for a planar detector
with hemispherical sky-coverage,
and the tangential surface
area $A$ of the detector, we arrive at
the rate of UCL and UAS events:
\begin{equation}
R_{_{\tau({\rm UAS})}} =  F_{\nu_\tau} \pi A
\int_0^{2R_\oplus}
\frac{l\,dl}{2\,R_\oplus^2}
\,P_{{\nu_\tau}\rightarrow{\tau({\rm UAS)}}}(l)\,.
\label{prob2}
\end{equation}
In Fig.~\ref{fig2} we show the number of expected UAS events per
incoming neutrino as a function of the neutrino cross section.
For comparison, we also show the number of expected HAS events per neutrino
that crosses a 250~km field of view, up to an altitude of 15~km.
It is clear that for the smaller values of the cross section, UAS events
will outnumber HAS events, and vice versa.

We give some examples of the UAS event rates expected from
a smaller neutrino cross section at $10^{20}$~eV, choosing 
$\sigma_{\nu N}=10^{-33}{\rm cm}^2$ as an example.
EUSO and OWL have shower-energy thresholds
$E_{\rm th}\sim 10^{19}$eV corresponding to curve 2 in Fig.~\ref{fig2},
and apertures $\sim 6\times 10^4$km$^2$ and
$3\times 10^5$km$^2$, respectively.
These detectors should observe $F_{20}$ and $7F_{20}$ UAS events per year,
respectively (not including duty cycle);
here $F_{20}$ is the incident neutrino flux at
and above $10^{20}$~eV in units of km$^{-2}$sr$^{-1}$yr$^{-1}$,
one-third of which are $\nu_\tau$'s.
Including showers from taus originating outside the
field of view, and direct tau events, increases these rates,
as does tilting space-based detectors
toward the horizon to maximize the
acceptance for events with smaller chord lengths
and to allow more atmospheric path length for tau decay.
The rates will also increase if $E_{\rm th}$ can be reduced.

Recently, it was pointed out that production of the lightest
supersymmetric particle (LSP) from annihilating
dark matter may present a detectable flux of LSP-CRs.~\cite{LSPflux}
A higgsino-dominated LSP may have a 
cross-section as small as $10^{-2}$ times the SM neutrino
cross-section.  Such a tiny cross-section is a perfect example 
of what can be discovered and measured
via either the UAS/HAS or the angular-dependence methods discussed here.

\section{Dispersion Relations: the High-Energy/Low-Energy Connection}
\label{sec:disperse}

A neutrino primary is not subject to the GZK losses
that limit cosmic nucleon propagation.
However, the extrapolated neutrino-nucleon cross-section is expected to be 
be $10^{-31}$~cm$^2$, about $10^{-6}$ too small to provide the neutrino 
with interactions high in our atmosphere.
To explain the production of the observed EE cosmic ray events
with neutrino primaries, some have postulated a new strong-interaction
for neutrinos above $\sim 10^{19}$ eV.
The idea of a strongly-interacting neutrino is not new,
but recent developments in field theory and in gravity 
have given new motivation to such a picture.
To mimic hadronically--induced air showers,
the new neutrino cross section must be of
hadronic strength,
$\sim 100$~mb, above $E_{\rm GZK}= 5\times 10^{19}\ \rm{eV}$.
Simple perturbative
calculations of single scalar or vector exchange
cannot provide an acceptably fast growth
of the cross-section with energy.~\cite{BGHpert}
However, the modern thoughts on large TeV-scale cross-sections are
much more imaginative.  A plethora of new states,
possibly growing exponentially in $s$ or $\sqrt{s}$,
is motivated by precocious unification, low-scale string theory,
and modes from additional space-dimensions accessible at
$\sqrt{s}\sim$~TeV.
Electroweak instantons are the most recent possibility.~\cite{EWinstant}

Direct limits on the EE neutrino cross-section\ are quite weak.
The vertical column density of our atmosphere is
$X_{\rm v}=1033\,{\rm g/cm}^2$.
In terms of neutrino MFP $\lambda_\nu$, this may be written
$X_{\rm v}/\lambda_\nu = \sigma_{\nu N}/1.6\,{\rm mb}$.
The horizontal slant depth $X_{\rm h}$ is 36 times larger,
leading to
$X_{\rm h}/\lambda_\nu = \sigma_{\nu N}/44\,\mu$b.
Since penetrating events are not observed above us or to our side,
the neutrinos must be interacting high in the atmosphere (large cross-section)
or not interacting at all (small cross-section).
Thus the cross-section\ range from $\sim 20\,\mu$b to $\sim1$~mb is 
excluded.
More quantitative analyses give similar results.~\cite{OST,AFGS}

A very interesting indirect limit on the EE neutrino cross-section 
is provided by dispersion relations.~\cite{GWdisp}
Dispersion relations are rigorous, nonperturbative,
and model-independent.  They limit the 
growth of the {\sl elastic} neutrino amplitude at low energy
due to any rising cross-section at higher energies.
If new physics dominates the neutrino total cross-section\ with
a value $\sigma^*$ above the lab energy $E^*$, then the dispersion relation
determines the real part of the new strong-interaction
elastic amplitude at lower energy $E$ to be
$\frac{1}{2\pi}\frac{E}{E^*}\sigma^*$.
Remarkably, significantly enhanced rates may occur
for elastic $\nu N$ scattering at an energy seven orders of magnitude
lower than the onset of a new
total cross--section.~\cite{GWdisp} 
Such anomalous ``low'' energy scattering may be observable in
the neutrino beams available at Fermilab and CERN,
and maybe in quasi-elastic $e^- p\rightarrow \nu_e n$ scattering at HERA.

How does this magic come about?
Assuming only that the scattering amplitude is analytic,
there results the following dispersion relation:~\cite{GWdisp} 
\begin{equation} {\rm Re}\ A_\pm (E)-{\rm Re}\
A_\pm(0)=\frac{E}{4\pi}\mathcal{P}\int_0^{\infty}\ dE^{\, \prime}\ \left(
\frac{\sigma^{\nu N}_{tot}(E^{\, \prime},\pm)}{E^{\, \prime}(E^{\, \prime}-E)}\ +\
\frac{\sigma^{\bar \nu N}_{tot}(E^{\, \prime},\pm)}{E^{\, \prime}(E^{\, \prime}+E)}
\right) \label{dispsubt} \end{equation}
where $A_\pm (E)$ are invariant $\nu$-$N$ amplitudes, labeled by
the nucleon helicity, and $\mathcal{P}$ denotes the principle
value of the integral. Suppose the new physics dominates the
neutrino-nucleon dispersion integral (\ref{dispsubt}) for
$E^{\, \prime}\ge E^*$ as hypothesized to explain the air
showers observed above the GZK limit. 
Assuming that
$\sigma^*$ is independent of helicity and energy, 
and obeys the Pomeranchuk theorem: 
$\sigma^{\nu N}_{tot}(E,\pm)-\sigma^{\bar \nu
N}_{tot}(E,\pm) \stackrel{\nu\rightarrow\infty}{\longrightarrow} 0$,
the real part of the amplitude at energy $E$ emerges:
\begin{equation}
{\rm Re}\ A_\pm(E)\simeq {\rm Re}\ A_\pm(0)
+ \frac{1}{2\pi}\frac{E}{E^*}\sigma^*\ \ .
\label{delta}
\end{equation}
This result cannot be obtained in perturbation theory! 
${\rm Re}\ A_\pm(0)$ is nothing but
the low energy limit of the weak interaction,
$\sim \frac{G_F}{2\sqrt{2}}$.  
From this, we may immediately write down
the ratio of the new amplitude to the SM amplitude:
\begin{equation}
\frac{{\rm Re}A(E)_{\rm new}}{{\rm Re}A(E)_{\rm SM}}
\simeq\left(\frac{E/100\ {\rm GeV}}{E^*/10^{18}\ \mbox{eV}}\right)\
\left(\frac{\sigma^*}{100\ \rm{mb}}\right)\ \ .
\label{r}
\end{equation}
It is clear from (\ref{r}), and striking, that order 100\% effects in
the real elastic amplitudes begin to appear already at energies seven orders of
magnitude below the full realization of the strong cross section.

A promising observable consequence is available from the elastic
cross section, obtained from the square of the elastic amplitude.
The result (\ref{dispsubt})
says that if the neutrino is strongly interacting at
$E^* \sim 10^{17.5}{\rm eV}$,
then a factor of ten anomalous rise
in the elastic cross--section is
occurring at 100~GeV, a neutrino energy
already available at Fermilab and CERN.
Since the anomalous elastic cross-section\ grows
quadratically with $E$, the anomalous event rate develops rapidly
for $E > 100$~GeV.
Thus, the event sample of a future underground/water/ice neutrino telescope
optimized for TeV neutrinos could conceivably contain {\em 1000} times more
elastic neutrino events than predicted by the SM;
and a telescope optimized for PeV neutrinos may contain $10^9$ more elastic
events.
Some of the wilder brane-world cross-sections proposed for the EE neutrino
are ruled out by this dispersion result.

There may be further tests of the strong--interaction hypothesis.
If the neutrino develops a strong-interaction\ at high energy,
do not the electron and the other charged-lepton $SU(2)$--doublet partners
of the neutrinos also develop a similar strong-interaction?
Is there new physics in the quasi-elastic $e^- p\rightarrow \bar{\nu}_e n$
scattering channel at HERA energies?
A possible enhancement in the quasi--elastic channel cannot
be deduced from dispersion relations. A separate calculation can be made,
however, if certain aspects of the new high-energy strong-interaction are
assumed.  This is presently under investigation.~\cite{GSWhera}

\section{Puzzles in the Extreme-Energy Cosmic Rays (EECRs)}
\label{sec:EECR}
The discoveries by the AGASA, Fly's Eye, Haverah Park,
and Yakutsk collaborations of air shower events with
energies above the GZK cutoff 
challenge the SM of particle physics and the hot
big-bang model of cosmology.
Not only is the mechanism of particle
acceleration for such EECRs controversial,
but also the propagation of EECRs over cosmic distances is problematic.
As has been mentioned, however,
the HiRes experiment does not confirm the prior AGASA rate
for events $\sim 10^{20}$~eV.  The situation will remain murky until the
Auger hybrid detector provides guidance.

\begin{figure}[th]
\centerline{\epsfxsize=13cm\epsfbox{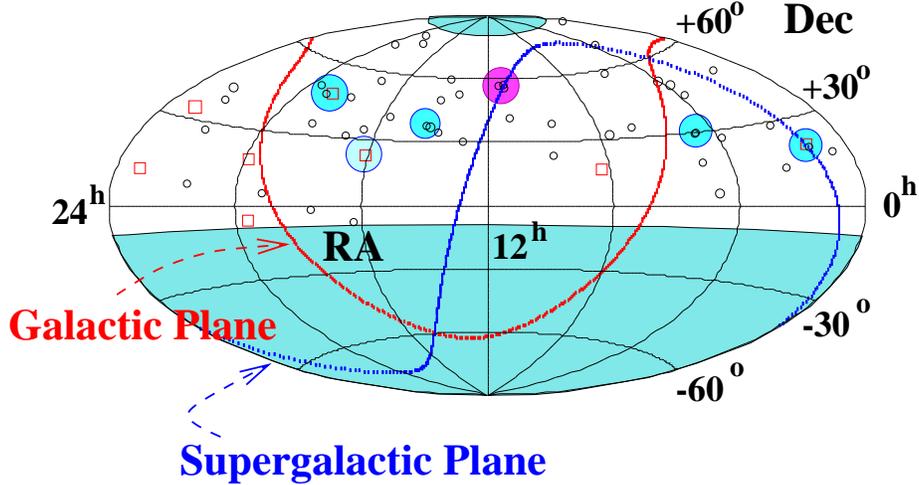}}
  \caption{AGASA sky-map of the arrival directions of the highest-energy 
cosmic rays (from the AGASA homepage).
The five light-shaded disks contain doublets,
and the one darker disk contains a triplet.
Disk solid angles are $\pi\,(\theta/2)^2$,
with cone opening-angle $\theta=2.5^\circ$.
} 
\label{fig:AGASAdirection}
\end{figure}

The famous Fly's Eye event occurred high in the
atmosphere, whereas the event rate expected in the SM for early development of
a neutrino--induced air shower is down from that of an electromagnetic or
hadronic interaction by six orders of magnitude. On the other hand,
Fig.\ \ref{fig:AGASAdirection} presents the AGASA 
evidence that the arrival directions of some of the
highest--energy primaries are paired.~\cite{AGASApairs}  
Furthermore, a recent analysis of the arrival directions of the super--GZK
events offers a tentative claim of a correlation with the directions of
BL-Lac quasars.~\cite{BLlac}   
This correlation, if validated with
future data, and the pairing data, argue for
propagating cosmic particles which are charge neutral, stable, 
and have a negligible magnetic moment.  
The neutrino emerges as the only candidate
among the known particles.

Several non-neutrino solutions
for the origin of the exceptional EECRs have been proposed.
Magnetic caustics have been proposed as a focusing
mini-lens bringing charged particles into pairs,
and it is conceivable that some new physics at 
high energy could stabilize the neutron.
With magnetic caustics one gets pairing but not pointing;
with stabilized neutrons, one gets pairing and pointing.
Generally, other proposed models are distinguishable 
from these and the neutrino scenario by the lack of pairing and pointing.

The unexpected degree of small-scale clustering observed in
the highest-energy AGASA cosmic ray day has motivated a derivation of
analytic formulas\cite{GWplets}
which estimate the probability of random cluster configurations.
The derived formulas offer a quick study of the strong potential of
HiRes, Auger, and EUSO
for deciding whether any observed clustering is meaningful or random.
For detailed comparison to data, this analytical approach cannot
compete with Monte Carlo simulations including experimental systematics.
Nevertheless, a recent Monte Carlo calculation\cite{Burgett}
of joint multiplet-probabilities 
shows reasonably good agreement with the analytic approach presented here.
(The same MC analysis also cross-correlates different energy bins to suggest 
that pairing is emerging only above $10^{19}$~eV.)
The derived formulas here do offer two advantages over Monte Carlo techniques:
(i) easy assessment of the significance of any observed clustering,
and most importantly,
(ii) an explicit dependence of cluster probabilities on the
chosen angular bin-size.

To derive the combinatoric formula for the probability
of various event distributions in angle, imagine that the sky coverage
consists
of a solid angle $\Omega$ divided into $N$ equal angular bins,
each with solid angle $\omega\simeq\pi\theta^2$~steradian.
Define each event distribution by specifying the partition of the $n$ total
events into a number $m_0$ of empty bins, a number $m_1$ of
single hits,a number $m_2$ double hits, etc.
The probability to obtain a given event topology is:
   \begin{equation}
   P(\{m_j\},n,N)=\frac{1}{N^n}\:
   \frac{N!}{m_0!\ m_1!\ m_2!\ m_3!\ldots}\:
   \frac {n!}{(0!)^{m_0}\ (1!)^{m_1}\ (2!)^{m_2}\ (3!)^{m_3}\ldots}\;.
   \label{eq:Hys}
   \end{equation}

The variables in the probability are not all independent.
The partitioning of events is related to the total number of events by
$\sum_{j=1} j\times m_j = n$,
and to the total number of bins by
$\sum_{j=0} m_j = N$.
Because of these constraints,
one infers that the process is not described by a simple multinomial
or Poisson probability distribution.
It is useful to use these constraints
to rewrite the probability (\ref{eq:Hys}) as
   \begin{equation}
   P(\{m_j\},n,N)=\frac{N!}{N^N}\,\frac{n!}{n^n}\,
   \prod_{j=0} \frac{(\overline{m_j})^{\,m_j}}{m_j!}\;,
  \label{eq:rewrite}
   \end{equation}
where
   \begin{equation}
   \overline{m_j}\equiv N\left(\frac{n}{N}\right)^j\frac{1}{j!}\;.
   \label{eq:meanm}
   \end{equation}
In the $n\ll N$~limit, $\overline{m_j}$
is expected to approximate the mean number of $j$-plets,
and eq. (\ref{eq:rewrite}) becomes roughly Poissonian.
As an approximate mean, $\overline{m_j}$ defined in eq.\ (\ref{eq:meanm})
provides a simple estimate of cluster probabilities due to chance
for the $n\ll N$~case.

Two large-number limits of interest are $N\gg n\gg 1$, and $n>N\gg 1$.
With bin numbers typically $\sim 10^3$, the first limit applies to the
AGASA, HiRes, and Auger experiments; the second limit
becomes relevant for the EUSO experiment.
Valid when $N\gg n\gg 1$, one has:
   \begin{equation}
P(\{m_i\},n,N)\approx \mathcal{P} \left[ \prod_{j=2}
\frac{(\overline{m_j})^{m_j}}{m_j!} e^{-\overline{m_j}\,r^j (j-2)!} \right]\,,
\label{eq:largeN}
   \end{equation}
where $r\equiv (N-m_0)/n\approx 1$, and the prefactor (nearly unity)
$\mathcal{P}$ is
\begin{equation} \mathcal{P}=e^{-(n-m_1)}\,\left(\frac{n}{m_1}\right)^{m_1
+\frac{1}{2}}\,. \end{equation}
The non-Poisson nature of Eq.\ (\ref{eq:largeN}) is reflected in the
factorials and powers of $r$ in the exponents,
and the deviation of the prefactor from unity.

In the case where $n>N\gg 1$, higher $j$-plets are common and
the distribution of clusters can be rather broad in $j$.
We may write $\overline{m_j}$ in the approximate form:
\begin{equation}
\overline{m_j}\approx \sqrt{\frac{N^3}{2\pi
en}}\,\left(\frac{en}{jN}\right)^{j+\frac{1}{2}}\,.
\label{eq:appm}
\end{equation}
Extremizing this expression with respect to $j$,
one learns that the most populated $j$-plet occurs near $j\sim n/N$.
Combining this result with the broad distribution expected for large $n/N$,
one expects clusters with $j$ up to ${\rm several}\times \frac{n}{N}$
to be common in the EUSO experiment.

\begin{figure}[th]
\centerline{\epsfxsize=12cm\epsfbox{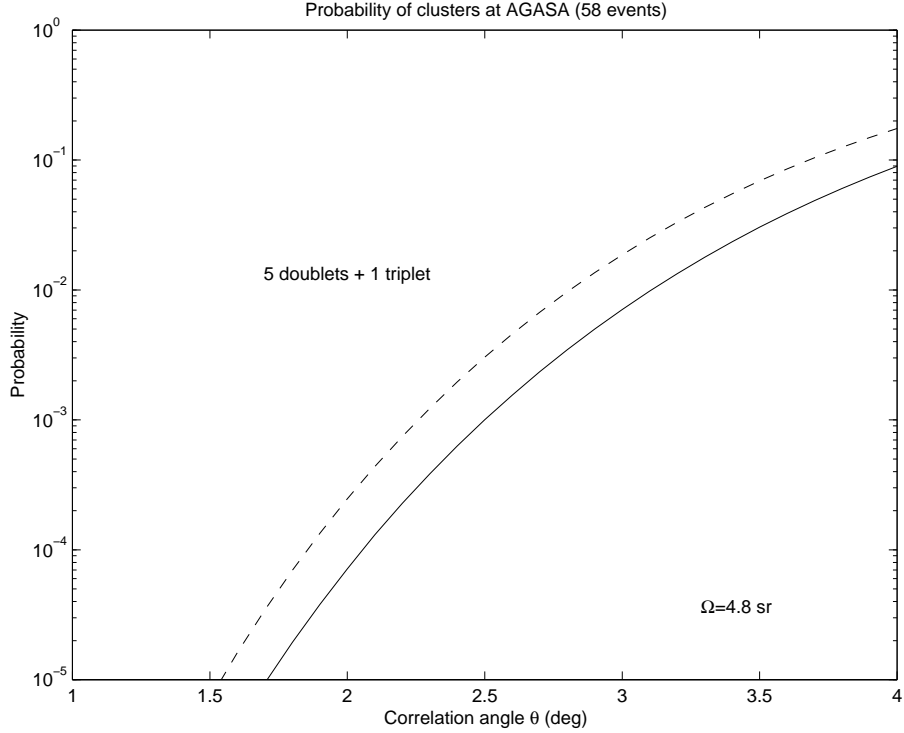}}
  \caption{Exact (solid) and Poissonian (dashed) inclusive
probabilities 
(from ref.\ 24) 
for five doublets and one triplet in the 58-event AGASA sample.}
\label{fig:AGASAprob}
\end{figure}

Shown in Fig.\ \ref{fig:AGASAprob} is an assessment\cite{GW-AGASA}
of the AGASA-plets, five doublets plus a triplet,
obtained using formula (\ref{eq:largeN}).
Two features of the figure are noteworthy.
The first is the rather extreme sensitivity of the statistical
significance to the angular binning size.
AGASA claims a $2.5^\circ$ resolution, which puts the significance
of their clusters at $10^{-3}$.
If the resolution were $3^\circ$ ($2.0^\circ$),
the significance would be a factor of six weaker (fifteen stronger).
The second feature is the factor of a few error made when Poisson statistics
are blindly applied.  

The explicit dependence of random clustering probabilities
on angular bin-size presented here
may prove quite useful in the future.  If clustered events
originated from a common source and traveled without bending,
then the experimental angular resolution is the optimal bin-size.
On the other hand, if clustering results from
magnetic focusing, then the angular size of magnetic caustics
may be the relevant bin-size.
If clustering results from density fluctuations in the Galactic halo,
then the angular size of these fluctuations may be
the optimal bin-size.
Since photons are not bent
by magnetic fields whereas protons are bent,
the optimal bin-size for photon-initiated
events is likely smaller than that for proton-initiated events.
The analytic formulas are easily applied to any
chosen angular bin-size.

\section{Z-bursts}
\label{sec:Zbursts}
\begin{figure}[th]
\centerline{\epsfxsize=9cm\epsfbox{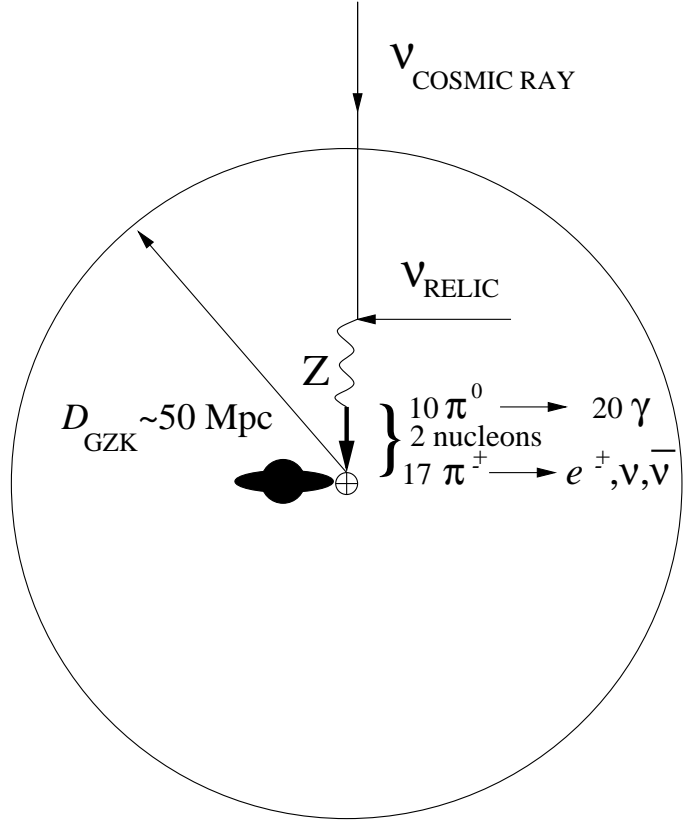}} \caption{Schematic
  diagram showing the production of a Z-burst resulting from the
  resonant annihilation of a cosmic ray neutrino on a relic
  (anti)neutrino.  If the Z-burst occurs within the GZK zone ($\sim$
  50 to 100 Mpc) and is directed toward the earth, then photons and
  nucleons with energy above the GZK cutoff may arrive at earth and
  initiate super-GZK air-showers}
\label{fig:Zburst}
\end{figure}

A rather conservative and economical solution proposed to solve the
super-GZK mysteries is the Z-burst mechanism.  Here, EECR
neutrinos scattering resonantly on the cosmic neutrino background (CNB)
predicted by Standard Cosmology, to produce Z-bosons.~\cite{Zburst} 
These Z-bosons in turn decay to produce a highly boosted ``Z-burst'',
containing on average twenty photons and two nucleons above $E_{\rm GZK}$
(see Fig.~\ref{fig:Zburst}).
The photons and nucleons from Z-bursts produced within 50 to 100 Mpc
of earth can reach earth with enough energy to initiate the
air-showers observed at $\sim 10^{20}$~eV.
The energy of the neutrino annihilating at the peak of the Z-pole is
\begin{equation}
E_{\nu_j}^R=\frac{M_Z^2}{2 m_j}=4\,(eV/m_j)\,{\rm ZeV}\,.
\label{Zburst}
\end{equation}
The resonant-energy width is narrow,
reflecting the narrow width of the Z-boson: at FWHM
$\Delta E_R/E_R \sim\Gamma_Z/M_Z = 3\%$.
The mean energies of the $\sim 2$~baryons and $\sim 20$~photons
produced in the Z decay are easily estimated,
when the Z-burst energy is averaged over the mean multiplicity
of 30 secondaries in Z-decay.
The photon energy is reduced by an additional factor of 2
to account for their origin in two-body $\pi^0$ decay.

In the simplest approximation, the spectrum of arriving nucleons is
\begin{equation}
\frac{dN}{dE} \sim \frac{1}{D^2}\times\frac{dN}{dD}
\times\frac{dD}{dE}\;\;\propto\;\;E^{-1}
\label{Zspectrum}
\end{equation}
from sources uniformly distributed out to
\begin{equation}
D_{\rm GZK}\sim \lambda\,\frac{\ln\left(\frac{N\,E_{\rm GZK}}{E^R}\right)}{\ln (1-f)}\,,
\label{Dgzk}
\end{equation}
with a pileup at 
$E_{\rm GZK}$ resulting from all primaries originating 
beyond this distance.  
The $E^{-1}$ spectrum extends from $E_{\rm GZK}$ out to the maximum 
nucleon energy $\sim E^R/30\sim 10^{21}(\frac{0.1{\rm eV}}{m_\nu})$~eV.
More realistic simulations including energy-loss processes, cosmic 
expansion, and boosted Z-boson fragmentation functions 
give a softer spectrum, but
a characteristic feature of the Z-burst mechanism remains 
that the super-GZK spectrum is considerably harder 
than the sub-GZK spectrum having power law index -2.7.

The necessary conditions for the viability of this
model are then,
a neutrino mass scale of the order 0.1~to~1~eV, and 
a sufficient flux of neutrinos at 
$\mathrel{\vcenter{\hbox{$>$}\nointerlineskip\hbox{$\sim$}}} 10^{21}$
~eV.~\cite{Zburst} 
The second condition seems challenging,
while the first is quite natural in view of the recent 
neutrino oscillation data.
Fits to atmospheric neutrino data yield
$m_\nu\ge\sqrt{\delta m^2_{\rm atm}}\sim 0.05$~eV.
A recent estimate\cite{2dFmass}
of the total neutrino mass in the Universe,
based on the distribution of large-scale structures,
is $\sum_j m_j \sim 2$~eV.
Most recently, the WMAP collaboration ambitiously 
combined their new CMB data with  
matter-distribution spectra to deduce\cite{WMAP} 
$\sum_j m_j \lsim 0.71$~eV.
However, omission of the suspect Lyman-alpha data 
from the analysis returns one to 
$\sum_j m_j \lsim 1$~eV, with dependence on priors\cite{noLyalpha}.
And so it appears that 
the neutrino mass is squeezed to lie within just the
0.1 to 1.0~eV range most beneficial to the Z-burst model!~\cite{BRSW}

What is not known is whether Nature has provided the large 
neutrino flux at energy $E^R$ to allow an appreciable event rate in 
future EECR detectors.
It is conceivable, although unlikely, that the flux is 
so large that present EECR events are initiated by Z-bursts.
A recent analysis\cite{FKR} of this possibility
gave a best fit with 
$m_{\nu}=0.26^{+0.20}_{-0.14}$~eV,
nicely consistent with the WMAP bound.
Another analysis\cite{Gelmini} fits the EECR spectrum down to the ankle
with Z-burst generated events and a neutrino mass of 
$\sqrt{\delta m^2_{\rm atm}}\sim 0.07$~eV,
again in accord with the WMAP bound.
The flux requirements for the Z-burst mechanism can be ameliorated
if there is an overdensity of relic neutrinos, as would happen 
if (i) there was a significant chemical potential, 
or (ii) neutrinos were massive enough to cluster in ``local''
structures such as the Galactic SuperCluster.  Large chemical potentials
have been ruled out recently\cite{BellBeacom}, 
and this exclusion is confirmed by the WMAP data.  
Local clustering has been studied,\cite{MaSingh}
with the conclusion being that a significant overdensity on the 
SuperCluster scale requires a neutrino mass in excess of 0.3~eV.
Such a mass is marginally allowed by the new WMAP limit.

The large neutrino flux required at $E^R$ 
to explain the AGASA super-GZK
event rate with the Z-burst mechanism has engendered much debate.
While the direct neutrino flux limits do not preclude 
such a large flux,\cite{AFGS} 
limits on the associated diffuse gamma-ray flux disfavor most
source mechanisms.  A wide range of possible 
sources have been critically reviewed recently.~\cite{NuSources,Venya} 
Even the so-called hidden sources (which emit only neutrinos)
seem problematical.~\cite{GTT} 
On the other hand, new possibilities have emerged
for production of neutrino fluxes.
Wake-fields in plasmas\cite{wakefield} 
may accelerate particles to energies
of $10^{23}$~eV, and a special class of blazars may 
beam neutrinos at us.~\cite{NS} 
``Mirror matter'' topological defects may decay 
to mirror neutrinos which then may oscillate into 
active neutrinos\cite{Venya,mirrormatter}.
Recombination of the strong magnetic fields
surrounding black holes offers another acceleration mechanism that 
is still in the infancy of its exploration.~\cite{recombo}

\section{Summary and Acknowledgments}
The next ten years will exploit neutrino eyes to see the extreme Universe.
Theory and experiment are rapidly converging to usher in the 
neutrino window to astronomy, and the astronomy window into 
neutrino physics.  Fruitful discoveries, probably beyond
anything anticipated today, await us.

I acknowledge my collaborators for the joint work I have presented here.
They are J. Beacom, N. Bell, G. Bhattacharyya, H. Goldberg, D. Hooper, 
A. Kusenko, H. P\"as, S. Pakvasa, and L. Song.
This work is supported by the U.S. Department of Energy 
grant no.\ DE-FG05-85ER40226, and by the Kavli ITP visitor program.

\end{document}